# Two examples of the relation between the contemporary science and Plato

Elio Antonello
*INAF-Osservatorio Astronomico di Brera*
elio.antonello@brera.inaf.it

**Abstract.** The philosopher Plato is remembered even today by scientists, and his writings are still inspiring the scientific research. In the present short note (intended essentially for public outreach) two examples are briefly illustrated: 1) the European space project that bears his name, dedicated to the discovery of exoplanets; 2) the discussion about 'platonism' in contemporary physics.

## 1. Introduction

Plato is often quoted by present day scholars, both scientists and humanists, and platonism can be considered a 'contemporary' view. It is the view that "there exist such things as abstract objects — where an abstract object is an object that does not exist in space or time and which is therefore entirely non-physical and non-mental" (Balaguer 2016). The modern platonism has been endorsed by numerous philosophers, mainly in relation to the foundations of logic and mathematics (see e.g. Hacking 2014).

In this note I will present briefly an example of the cultural influence of Plato on the contemporary astronomical research, that is the one regarding the exoplanets, and, by the way, I will recall also some implications of the search for Earth-like planets. Then I will illustrate an example of platonism in contemporary physical ideas, and I will mention also some different points of view.

## 2. A space project

In 2006 there was an ESA Call for Proposals of Space Missions, and an European team of scientists responded proposing the project of a space satellite dedicated to the search for exoplanets, using the observing technique of the transits of the planets in front of the parent star. Owing to such a transit, the luminosity of the star decreases slightly; it is a very small eclipse: the smaller the planet, the smaller the eclipse. The turbulence of the Earth's atmosphere introduces significant disturbances in the photometric measurements; therefore, in general, the long term observations for detecting and confirming a transit must be performed from space.

The European team adopted the name PLATO for the project, an acronym whose meaning is PLAnetary Transits and Oscillations of stars. The motivation for this choice was that, although Plato did not give significant contribution to astronomy by himself, his mathematical conception of Nature is on the foundation of the Western scientific thought. In the website of the project PLATO 2.0[1], one can read the question that, according to Sosigenes quoted by Simplicius of Cilicia (6th century CE), Plato asked: "By the assumption of what uniform and orderly motions can the apparent motions of the planets be accounted for?" (Burkert 1972, p. 329)[2].

---
[1] http://www.oact.inaf.it/plato/PPLC/Home.html  (August 2016).

The scientific purposes of the mission may be summarized as follows (Rauer et al. 2014): 1) To reveal the interior of planets and stars, by analysing the dynamics of the orbit and the very small variations of the stellar luminosity (stellar seismology). 2) To detect planets over the most part of the sky, and in particular those in the habitable zone around the parent stars, by using 34 small aperture optical telescopes that provide a wide field-of-view. 3) To put constraints on the planet formation and evolution, by making a comparison with the predictions of the models of the dynamical evolution of planetary systems. 4) To provide accurate ages of planetary systems, as it had been possible for our Solar system. 5) Finally, to provide the targets for the study of the atmosphere of those planets, by means of spectroscopy. A follow-up program for the stellar radial velocity measurements will permit the accurate determination of the mass of the planets.

As for August 2016, more than 3000 extrasolar planets have been discovered and confirmed[3]. The observations were performed with ground based telescopes mainly for radial velocity measurements, and with space instruments for the transits. For most of these planets it was possible to determine only one of the their fundamental parameters directly, radius or mass. Most of them are similar to Jupiter and Saturn, and the range of their distances from the parent star is very large, from one hundredth to many tens astronomical units. Very few Earth-size planets in systems with a similar parent star to the Sun and at a distance of about one astronomical unit (Earth-Sun distance) have been discovered; they are located too far from us and hence their detailed studies with telescopes discussed in the next Section are not possible. However, the researchers think that rocky planets like Earth should be very common around solar-type stars.

Most of the known transits were observed by the two space missions KEPLER and COROT; so one could wonder why another satellite? COROT and KEPLER observed small regions of the sky, with many relatively faint stars. PLATO will cover practically 50 per cent of the sky, and it will observe a large fraction of the brightest and nearest stars, increasing the probability to detect planets with a similar size to that of Earth, and located in the habitable zone. PLATO has been selected by ESA for the M3 launch opportunity (in 2022-2024).

**3. Are we alone?**

A terrestrial planet located in the habitable zone is not too close to and not too far from the parent star, so if water is present it is liquid. If such a terrestrial planet were detected not too far from us (maybe the one orbiting Proxima Centauri; Anglada-Escudé et al. 2016), the next step would be to observe its atmospheric features. This requires the largest telescopes from ground and from space. For example, the ESO - European Southern Observatory has planned the realization by the next decade of the E-ELT (European Extremely Large Telescope), the largest telescope in the world. According to ESO researchers, the E-ELT hopes to bring us one step closer to answering the question: are we alone? Clearly, this would represent a major breakthrough for humanity[4].

In principle, it is possible to detect the spectrum of the atmosphere of a planet in the infrared band, if the stellar mass is not too large and the planet is Jupiter-like. One can detect water, carbon dioxide, methane (see e.g. Seager, Deming 2010). But what about the atmosphere of a terrestrial planet? The dream of the scientists is to detect a spectrum with the signature of life. The infrared spectrum of the Earth's atmosphere observed from space shows

---

[2] *Simplicii in Aristotelis De Caelo Commentaria*, ed. I.L. Heiberg, 1894, p. 488. Plato rejected as 'blasphemous' the apparent irregularity of planetary motions (see e.g. Antonello 2016).
[3] http://exoplanetarchive.ipac.caltech.edu/, http://exoplanet.eu/
[4] https://www.eso.org/public/unitedkingdom/teles-instr/e-elt/e-elt_exo/ (August 2016)

the presence of water and carbon dioxide, but they are not indicators of life by themselves. The observed oxygen molecule would be the expected signature, since a large amount of oxygen could be reasonably explained only as a biological product. If we could go back in time and observe the spectrum of our atmosphere from space, it would be quite different from the present one. Today the air is composed mainly of nitrogen (78%), oxygen (about 21%) and small amounts of other gases. The formation of the Solar system and the Earth occurred about 4.5 billion years ago, and from then and for almost 2 billion years the oxygen content of Earth's atmosphere should have been negligible. It started to increase quickly about 2.33 billion years ago, reaching a value of just few per cent (e.g. Kump 2008; Luo et al. 2016). Only after the Ediacaran, and the Cambrian explosion of multicellular life, about 500 million years ago, it reached probably a value above 10%, with a maximum (upper limit) value of about 33% during the Carboniferous (Kump 2008), about 300 million years ago. Presumably, the Earth life could have been detected 'from space' with some confidence 'only' during the last 500 million years, when there was a sufficient amount of oxygen. In other words, the possibility to detect oxygen in the atmosphere of another terrestrial planet will depend also on its age and its specific evolutionary history.

But what about intelligent life and civilization? Its evolution phase, that is, the last ten thousand years, is completely negligible in the history of Earth. Therefore the probability to detect civilization signs (assuming it were technically possible) in a generic extrasolar terrestrial planet located in the habitable zone, with sufficient oxygen in the atmosphere, could be negligible. The probability would increase if it were a very long-lived civilization.

It may be worth to adopt an astronomical perspective to study the evolution of the Earth and of the mankind: i.e. to consider the Earth just one of the planets in the Solar system or an exoplanet, and to consider the scientists as people that are observing the Earth from 'outside', in a certain sense being not earthlings, as I suggested recently (Antonello 2013). Hence, if we adopt an astronomical perspective, and we do not think in an 'anthropocentric' way, the question "Are we alone?" posed by the earthlings, although very fascinating, could be a little misleading, at least for the present level of technological development of the instruments.

Other authors have different opinions, and there are many interesting papers related, for example, to the SETI project, the Search for Extraterrestrial Intelligence. Gerig et al. (2013) discussed the "universal doomsday", that is, the possibility of the existence of long-lived civilization and that we are living in one of them. Those authors wrote: "given a sufficiently large universe, numerous civilizations almost surely exist". This sentence is not in conflict with our remark about the present limitations. A "sufficiently large universe" means at least a hundred billion galaxies, each with a hundred billion stars, that is a total of at least $10^{22}$ stars, while we are talking about only $10^6$ stars that presumably could be surveyed for exoplanets by our current instruments and by those of the next future.

**4. Platonism in contemporary physics**

An example of how Plato is inspiring physicists today is represented by Penrose (2005). In *The Road to Reality* (Chapter 1), he recalls the influence of the Pythagoreans on the progress of human thought. With their introduction of the mathematical proof, it was possible to make assertion of an unassailable nature, so that they would hold just true even today. About a century and half after Pythagoras, Plato made it clear that the mathematical propositions referred not to actual physical objects but to idealized entities, that inhabited not the physical world, but that of mathematical forms. Is Plato's mathematical world 'real'? Many people, including philosophers, might regard such a world as a product of our imagination. Plato's viewpoint however is a valued one, since he distinguishes the precise mathematical entities from the approximations that we see around us in the world of physical things. The mathematical world possesses an objectivity that transcends mere opinion. Platonic existence refers to the existence of an objective external standard that is not dependent upon our individual opinions nor upon our particular culture. Platonic existence is simply a matter of objectivity, and, accordingly, should certainly not be viewed as something 'mystical' or 'unscientific', despite the fact that some people regard it that way.

Mathematics is concerned with the particular ideal of Truth. Plato himself would have insisted that there are two other fundamental absolute ideals, namely that of the Beautiful and of the Good. Aesthetic criteria are fundamental to the development of mathematical ideas, providing both the drive towards discovery and a powerful guide to truth. An important element in the mathematician's common conviction that an external Platonic world actually has an existence independent of ourselves comes from the extraordinary unexpected hidden beauty that the ideas themselves so frequently reveal.

Mathematical existence is different not only from physical existence but also from an existence that is assigned by our mental perceptions. Yet there is a deep and mysterious connection with each of those other two forms of existence: the physical and the mental. There may be a sense in which the three worlds – the Platonic mathematical, the physical and the mental – are not separate at all, but merely reflect, individually, aspects of a deeper truth about the world as a whole of which we have little conception at the present time.

Modern physicists describe things in terms of mathematical models; it is as though they seek to find 'reality' within the Platonic world of mathematical ideas. It is undoubtedly the case that the more deeply we probe Nature's secrets, the more profoundly we are driven into Plato's world of mathematical ideals as we seek our understanding. At present, we cannot explain this mystery.

Of course, other scientists have different opinions from those of Penrose. Hacking (2014) discusses in some detail different philosophical attitudes from platonism concerning mathematics, such as intuitionism, formalism, nominalism, structuralism[5]. For example, quoting Connes, he remarks that the Platonist attitude consists in saying that there exists a mathematical reality that precedes the elaboration of concepts (mathematicians are discoverers), and the formalist or structuralist one consists in treating mathematics as a system of logical deductions obtained at the interior of a language starting from axioms, denying the ontological character of mathematical reality (mathematicians are inventors). A firm anti-Platonist would just adopt the known maxim: don't ask for the meaning, ask for the use (Hacking 2014, p. 203, p. 216).

Smolin (2013) recalls the orthodoxy among theoretical physics, that considers the distinction between past, present and future as an 'illusion'; according to this view, truth is timeless and somehow outside the universe. He remarks that this was the essential of Plato's thought, since, as mentioned in *Meno*, all discovery is merely recollection (Smolin 2013, p. XV). Now, however, Smolin is proposing that time is not an 'illusion' but it is 'real'. The radical suggestion is the insistence on the reality of the present moment, and the principle that all that is real is so in the present moment. According to Smolin, the physical laws can change in time. "To the extent that this is a fruitful idea, physics can no longer be understood as the search for a precisely identical mathematical double of the universe. That dream must be seen now as a metaphysical fantasy that may have inspired generations of theorists but is now blocking the path to further progress" (Smolin 2013, p. 251). That is, there would be no mathematical world by its own, and the mathematical description would be a mere approximation to the reality. The radical position about time, however, seems to show philosophical weak points, according to the critique of the 'parmenidean' Price (2013): the philosopher feels "sadness for the unnecessity of Smolin's campaign", and for the fact that his own 'team philosophy' has not succeeded in making the things clear.

---

[5] See *Stanford Encyclopedia of Philosophy* for various entries; http://plato.stanford.edu/contents.html.

**5. A story**

Let me conclude by mentioning the *Prologue* of Penrose's book, where he tells a story related to Thera (Santorini, Greece). It begins with the destructive earthquake and tsunami following the Thera eruption, about 1600 BC, and Penrose imagined some people after the disaster, asking themselves what deep forces control the behaviour of the world. One thousand years later, about the 6$^{th}$ century BC, the great story of the catastrophe that destroyed an ancient peaceful civilization had been handed down from father to son, along with the same question. One clear night, a man looked up at the heavens; he puzzled over why the Gods had not organized the stars in a more appropriate way? Then an odd thought overtook him: do not seek for reasons in the specific patterns of stars; look, instead, for a deeper universal order in the way that things behave. The man then took to the sea. He found his way to Croton, where the sage and his brotherhood were in search of truth; the name of the sage was Pythagoras (Penrose 2005, pp. 1-5).

Although the 'man that looked up at the heavens' is just a story written by Penrose, I would consider it a further support to the view that the human knowledge began with the sky-gazing, a view that was proposed by Plato, in *Timaeus* (47a, b). It may be worth to recall once more this view about the possible common origin of both scientific and humanistic culture. Today, with the separation of culture in disciplines and sub-disciplines, such as in the Universities with the "perennial issue" of "the departmental structure that keeps researchers mentally and physically separated" (editorial in *Nature*, 514, p. 287; 16 October 2014), the sky-gazing could be considered the childhood home where they could meet again.